\documentclass[useAMS,usenatbib]{mn2e}
\usepackage{amsmath,amstext,amsgen,amsbsy,amsopn,amsfonts,theorem}

\usepackage[pdftex]{graphicx}
\usepackage{xspace}
\usepackage{amsmath}
\usepackage{hyperref}
\usepackage{framed}
\usepackage{txfonts}
\usepackage{epstopdf}
\usepackage{gensymb}

\usepackage{subcaption}

\usepackage[normalem]{ulem}

\makeatletter
\def\mr@ignsp#1 {\ifx\:#1\@empty\else #1\expandafter\mr@ignsp\fi}%
\newcommand{\multiref}[1]{\begingroup
\xdef\mr@no@sparg{\expandafter\mr@ignsp#1 \: }%
\def\mr@comma{}%
\@for\mr@refs:=\mr@no@sparg\do{\mr@comma\def\mr@comma{,}\ref{\mr@refs}}%
\endgroup}
\makeatother

\newcommand{\hypref}[2]{\ifx\href\asklfhas #2\else\href{#1}{#2}\fi}
\newcommand{\Secref}[1]{Section~\multiref{#1}}
\newcommand{\secref}[1]{Sec.~\multiref{#1}}

\newcommand{\Tabref}[1]{Table~\multiref{#1}}
\newcommand{\tabref}[1]{Tab.~\multiref{#1}}
\newcommand{\Figref}[1]{Figure~\multiref{#1}}
\newcommand{\figref}[1]{Fig.~\multiref{#1}}
\renewcommand{\eqref}[1]{(\multiref{#1})}

\title[Satellite census and LMC mass]{Limit on the LMC mass from a census of its satellites}

\author[D. Erkal and V. A. Belokurov]
  { Denis Erkal$^{1}$\thanks{d.erkal@surrey.ac.uk} \& Vasily A. Belokurov$^{2}$ \\
  $^1$Department of Physics, University of Surrey, Guildford GU2 7XH, UK \\
  $^2$Institute of Astronomy, University of Cambridge, Madingley Road, CB3 0HA, Cambridge, UK
   }

\begin{document}

\label{firstpage}

\maketitle

\begin{abstract}
We study the orbits of ultra-faint dwarf galaxies in the combined
presence of the Milky Way and LMC and we find 6 dwarfs which were
likely accreted with the LMC (Car 2, Car 3, Hor 1, Hyi 1, Phe 2, Ret
2), in addition to the SMC, representing strong evidence of dwarf
galaxy group infall. This procedure depends on the
gravitational pull of the LMC, thus allowing us to place a lower bound
on the Cloud's mass of $M_{\rm LMC} > 1.24\times10^{11} M_\odot$. This
mass estimate is validated by applying the technique to a cosmological
zoom-in simulation of a Milky Way-like galaxy with an LMC analogue
where we find that while this lower bound may be overestimated, it
will improve in the future with smaller observational errors. We apply
this technique to dwarf galaxies lacking radial velocities and find
that Eri 3 has a broad range of radial velocities for which it has a
significant chance ($> 0.4$) of having being bound to the Cloud. We
study the non-Magellanic classical satellites and find that Fornax has
an appreciable probability of being an LMC satellite if the LMC is
sufficiently massive. In addition, we explore how the orbits of the
Milky Way satellites change in the presence of the LMC and find a
significant change for several objects. Finally, we find that the
LMC satellites are slightly smaller than the Milky Way satellites at a
fixed luminosity, possibly due to the different tidal environments
they have experienced.
\end{abstract}

\begin{keywords}
 Galaxy: kinematics and dynamics, Galaxy: evolution, galaxies: Magellanic Clouds
\end{keywords}

\section{Introduction}

The hierarchical structure formation paradigm states that the smallest
objects collapse first, and then merge to create larger and larger
systems \citep{white-rees_1978}. A natural consequence of this is that
galaxies like our own are surrounded by a plethora of dwarf
satellites. In addition, it suggests that many of these dwarfs should
themselves have formed by accreting less massive objects and thus may
host their own, even smaller, galaxy companions.

The Dark Energy Survey (DES) mapped out a large fraction of the
Southern Galactic hemisphere which uncovered a plethora of dwarfs
close to the LMC
\citep[e.g.][]{koposov_des_sats,bechtol_etal_2015,drlica_wagner_etal_2016}. This
is in stark contrast to their distribution in the Northern Galactic
hemisphere where they show little clustering. Motivated by this
difference, \cite{jethwa_lmc_sats} modelled the expected distribution
of satellites that the LMC and SMC would bring to the Milky Way. Their
models found an excess of satellites was expected near the Magellanic
Clouds as well as along the past and future orbits of the LMC. Similary,
\cite{sales_etal_2017} used a cosmological simulation of a Milky Way-like
halo with an LMC analogue to predict the location and kinematics of the LMC satellites.

Based on the exquisite proper motions from \textit{Gaia} DR2,
\cite{kallivayalil_etal_2018} investigated which of the known
satellites belonged to the LMC. Their approach is based on comparing the
cosmological simulation of the LMC's infall from \cite{sales_etal_2017} with the satellite
population observed today. They found four satellites which were
consistent with their criteria for the expected population of the LMC
satellites. Given that this work was based on a single simulation, it
is unclear how the results would change if, for example, the
properties of the LMC or the Milky Way were altered.

In order to address this, this paper tackles the same problem with a
different technique. Instead of forward modelling the LMC satellites
to the present \citep[as in,
  e.g.][]{jethwa_lmc_sats,sales_etal_2017,kallivayalil_etal_2018}, we will rewind the
satellites from their present day positions and determine which
satellites were originally bound to the LMC. This approach has several
advantages. First, we can explore a large parameter space by varying
the properties of the LMC. Second, we can determine what LMC mass is
needed to bind each of the satellites in order to constrain the LMC
mass. Third, this method can be used to estimate the orbit of each
satellite relative to the Milky Way and the LMC. Finally, this independent technique
provides a useful check of the results in
\cite{jethwa_lmc_sats,kallivayalil_etal_2018}.

This rewinding technique was originally used in
\cite{kallivayalil_etal_2006b,kallivayalil_etal_2013} in order to
determine whether or not the LMC and SMC were accreted as a pair. It
is also similar to the method presented in
\cite{price-whelan_etal_2014}. An important difference introduced here is
that we also account for the motion of the Milky Way in response to
the LMC's infall. This effect was first highlighted in
\cite{gomez_et_al_2015}. Given the large LMC mass inferred from
abundance matching, the timing argument with Andromeda
\citep{penarrubia_lmc}, and from its deflection of the Orphan stream
\citep{orphan_modelling}, this effect is essential to include.

This paper is organised as follows. In \Secref{sec:mw_analysis} we
describe the rewinding method and then apply it to 32 ultra-faint
dwarf galaxies and the classical satellites. Additionally, in order to
test the technique, we apply it to a cosmological zoom-in simulation
in \Secref{sec:cosmo_tests}. We discuss some implications of these
results in \Secref{sec:discussion} and then conclude in
\Secref{sec:conclusion}.

\section{Method} \label{sec:mw_analysis}

\subsection{Satellite properties}

\textit{Gaia} DR2 \citep[][]{GaiaDR2} has delivered a plethora of data
on the Milky Way. Shortly after the data release, a number of works
measured the proper motions of dwarf galaxies around the Milky Way
\citep{simon_gaia_pms,fritz_gaia_pms,pace_li_pms}. Using these proper
motions, combined with radial velocities measured from other studies,
we have a sample of 25 ultra-faint dwarfs with 3d positions and 3d
velocities (see \tabref{tab:sats}). For the LMC properties, we use a
distance of $49.97 \pm 1.126$ kpc \citep{pietrzynski_lmc_dist}, a
radial velocity of $262.2\pm3.4$ km/s \citep{vandermarel_lmc_rv}, and
a proper motions of $\mu_\alpha^* = 1.91\pm0.02 \, {\rm mas/yr}$,
$\mu_\delta = 0.229\pm0.047 \, {\rm mas/yr}$
\citep{kallivayalil_etal_2013}.

\begin{figure}
\centering
\includegraphics[width=0.49\textwidth]{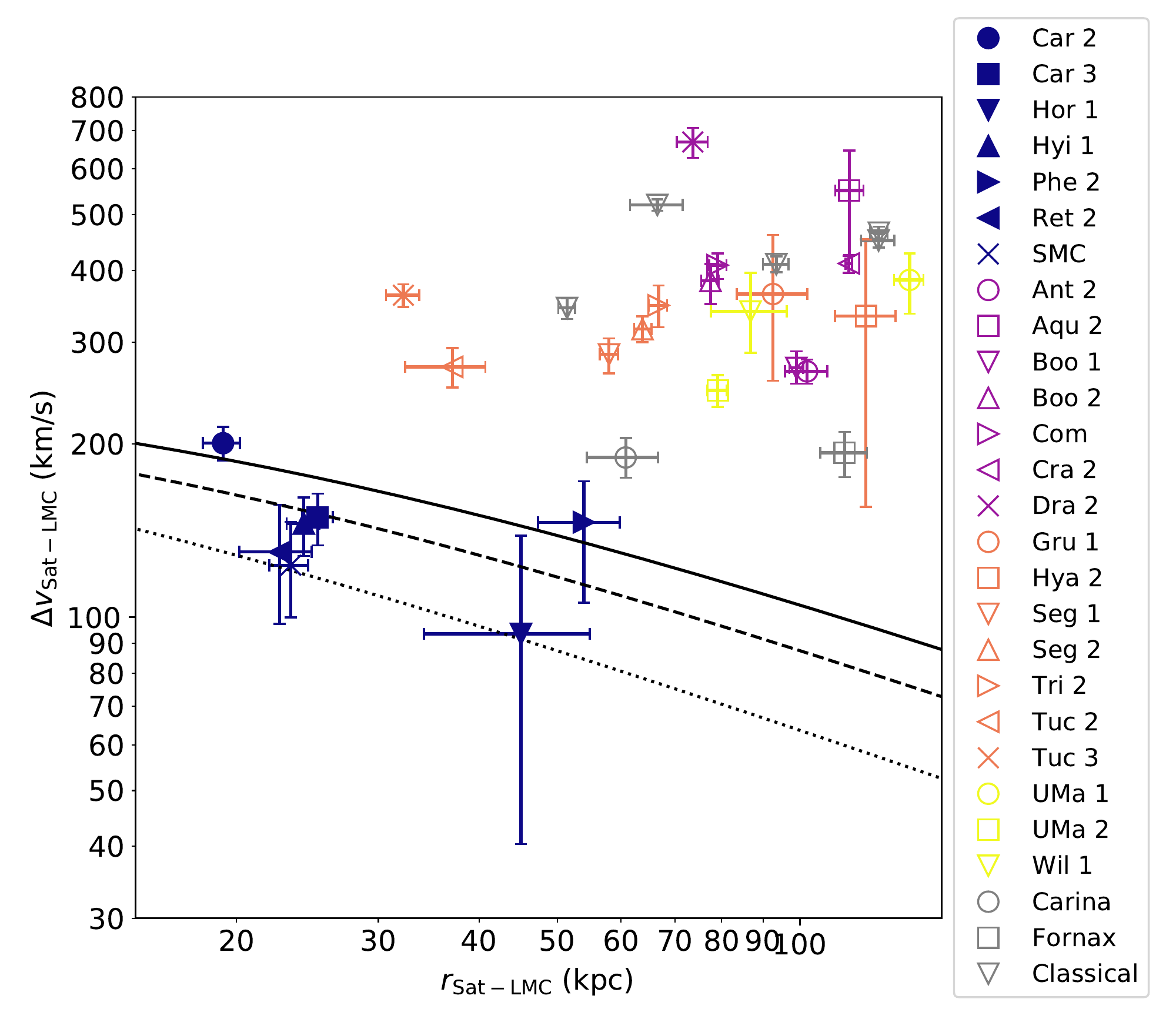}
\caption{Relative position and velocity between the LMC and ultra-faint dwarfs at the present. Somewhat unsurprisingly, the 7 satellites closest in position and velocity to the LMC (filled, dark blue symbols) are the ones which we find to be most likely to have fallen in with the LMC. Tuc 2 and Tuc 3, which are closer to the LMC than Hor 1, are not associated with the LMC. The classical dwarfs are shown in grey with two of the closest dwarfs in phase space, Carina and Fornax, labeled (see \protect\secref{sec:classical} for more details). Note that we limit this plot to dwarfs within 150 kpc of the LMC and thus do not show Eri 2, Her 1, Leo 1, Leo 2. We stress that the determination of whether each satellite is bound is determined by integrating their orbits backwards in time and not just using their present day position and velocity. The dotted, dashed, and solid lines show the escape velocity curves for a $5\times 10^{10}M_\odot$, $10\times 10^{10}M_\odot$, $15\times 10^{10}M_\odot$ LMC respectively with scale radii set as described in the text.}  \label{fig:vvel_rrel}
\end{figure}

In \Figref{fig:vvel_rrel} we show the relative position and velocity of these satellites with respect to the LMC. The curves show the escape velocity curve assuming different masses for the LMC. In each case, the LMC is treated as a Hernquist profile \citep{hernquist_1990} with a scale radius such that the circular velocity at 8.7 kpc matches the observed value of 91.7 km/s \citep{vandermarel_lmc}. Interestingly, this figure shows that there is a population of satellites which are close in both position and velocity to the LMC. We stress that we do not use the present day position in phase space to determine whether each satellite belongs to the LMC but rather rewind the satellites in time as described below.

In addition, there are 7 dwarfs which only have proper motions but no
radial velocity measurements. For these dwarfs, we will sample over
the possible radial velocities to determine whether there exist any
radial velocity for which they could be associated with the
LMC. Finally, we also repeat our analysis on the classical Milky Way
satellites.


\subsection{Probability of being an LMC satellite} \label{sec:lmc_prob}

In order to determine whether each satellite was originally associated
with the LMC, we Monte Carlo sample its $6d$ phase space position, as
well as the LMC's $6d$ position, and rewind them for 5 Gyr. The energy
of the satellite relative to the LMC is then computed to determine
whether the satellite was energetically bound. At this time, the LMC
is typically quite distant from the Milky Way and thus this also
selects satellites which were close to the LMC at this time. This
procedure is repeated 10,000 times for each satellite and LMC mass
combination. During this rewinding, we include the dynamical friction
of the Milky Way on the LMC using the prescription in
\cite{jethwa_lmc_sats}. We also account for the motion of the Milky
Way in response to the LMC which can move it a significant amount
\citep[e.g.][]{gomez_et_al_2015,orphan_modelling}. Each dwarf is
treated as a massless tracer under the combined influence of the Milky
Way and LMC.

We consider a grid of LMC masses: $2,5,10,15,20,25,30\times10^{10} M_\odot$. This lower limit is motivated by the mass constraint within $8.7$ kpc from \cite{vandermarel_lmc} and the upper limit is motivated by the LMC mass inferred in \cite{penarrubia_lmc} based on the nearby Hubble flow and the timing argument with Andromeda. The LMC is modelled as a Hernquist profile \citep{hernquist_1990} with a scale radius such that the circular velocity is $91.7$ km/s at 8.7 kpc as observed by \cite{vandermarel_lmc}. 

\begin{figure}
\centering
\includegraphics[width=0.49\textwidth]{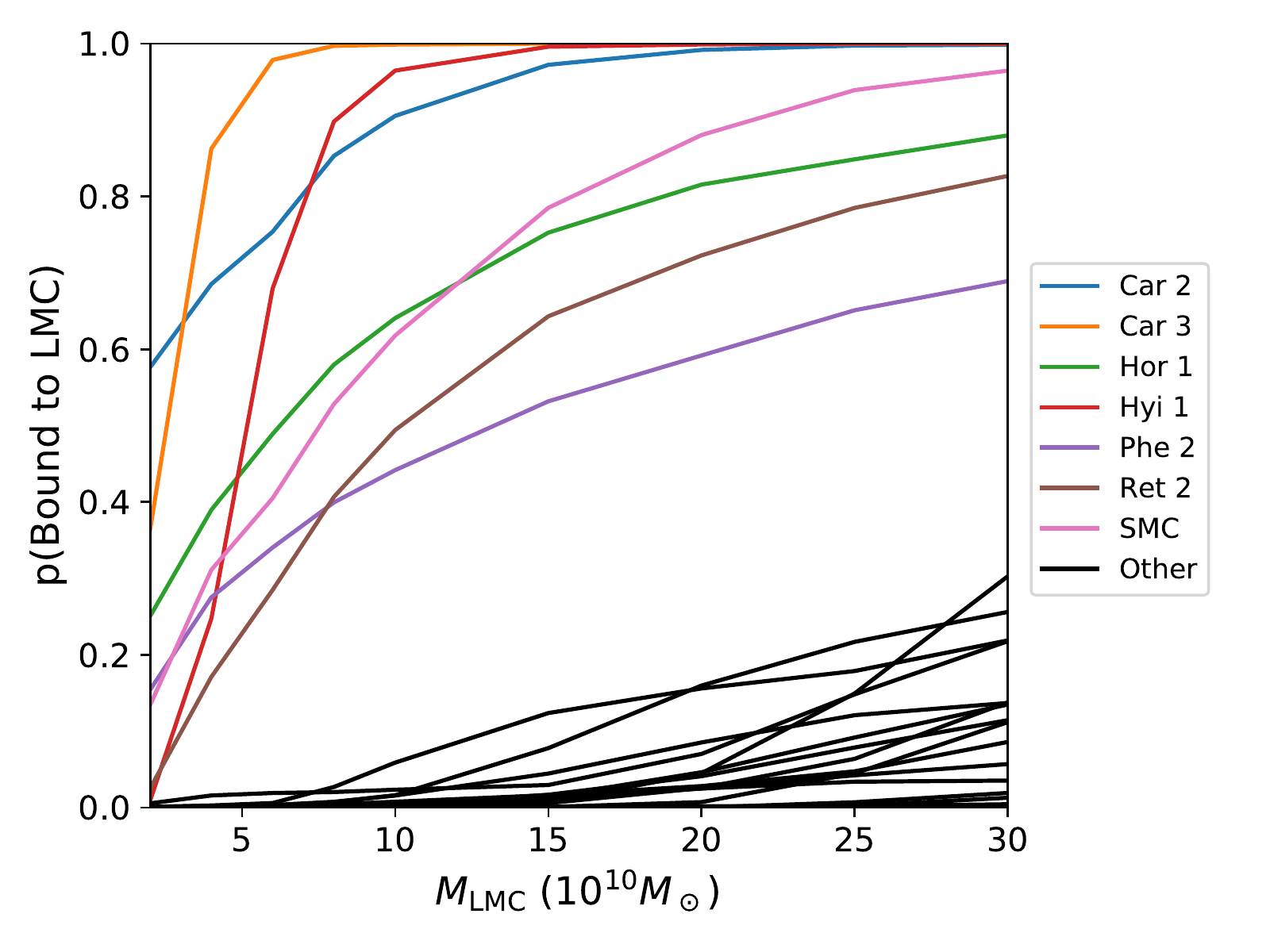}
\caption{Probability of being bound the LMC 5 Gyr ago as a function of satellite mass for the ultra-faint dwarfs. For each LMC mass, we Monte Carlo sampled the satellite's $6d$ phase space position (as well as the LMC's) and integrate the orbit backwards to determine whether the satellite is bound. As the LMC's mass is increased, we find that the satellites are more likely to be bound. The colored curves show the probabilities for the five likely bound satellites and the black curves show the probabilities for the remaining 12 satellites. Interestingly, four of the satellites (Hor 1, Car 2, Car 3, and SMC) have a non-zero probability of being bound even with the lowest mass we consider, $2\times 10^{10}M_\odot$ which is the observed mass of the LMC within 8.7 kpc \protect\cite{vandermarel_lmc}. The other 12 satellites, shown with black curves, have a negligible chance of being bound, with Tuc 2 and Tuc 3 reaching a bound probability of 3.3\% and 2.5\%, respectively, with an LMC mass of $3\times 10^{11} M_\odot$. } 
\label{fig:pbound}
\end{figure}

In \Figref{fig:pbound} we show the probability of being bound to the
LMC as a function of the LMC's mass. We find two distinct populations,
one which shows a significant probability of being bound even for
modest LMC masses. We postulate these to be LMC satellites. The second
population shows a low chance of being bound at low LMC masses which
increases with LMC mass but only reaches a modest probability even for
the most massive LMC we consider. We conclude that these are the Milky
Way satellites. Interestingly, some of the LMC satellites (Hor 1, Car
2, Car 3, and the SMC) have a non-zero probability of being bound even
with the lowest LMC mass considered suggesting that these sit deep in
the potential of the LMC. The binding probabilities for an LMC mass of
$1.5\times10^{11}M_\odot$ \citep[motivated by the results
  of][]{orphan_modelling} are given in \Tabref{tab:sats}.

\subsection{Mass of the LMC}

Next, we use the likely LMC satellites to estimate the mass of the
Cloud. This is done by assuming that each of the 7 dwarfs with a
significant chance of being bound is truly an LMC satellite. For each
object, we make 10,000 Monte Carlo realizations of its orbital
evolution in the presence of the LMC. For each realization, we begin
with an LMC mass of $2\times10^{10} M_\odot$ and rewind the satellite
and the LMC as described above to determine whether it is bound. If
the satellite is not bound, we increase the LMC mass by $10^9 M_\odot$
and repeat the rewinding procedure. This is repeated until we find an
LMC mass for which this particular realization of the satellite is
bound. We call this the minimum mass needed to bind the satellite
since it would also be bound for a more massive LMC. If the LMC mass
exceeds $3\times 10^{11} M_\odot$, we classify the satellite as
unbound and move onto the next realization. This process then yields a
distribution of the minimum LMC mass needed to bind the satellite.

\begin{figure}
\centering
\includegraphics[width=0.45\textwidth]{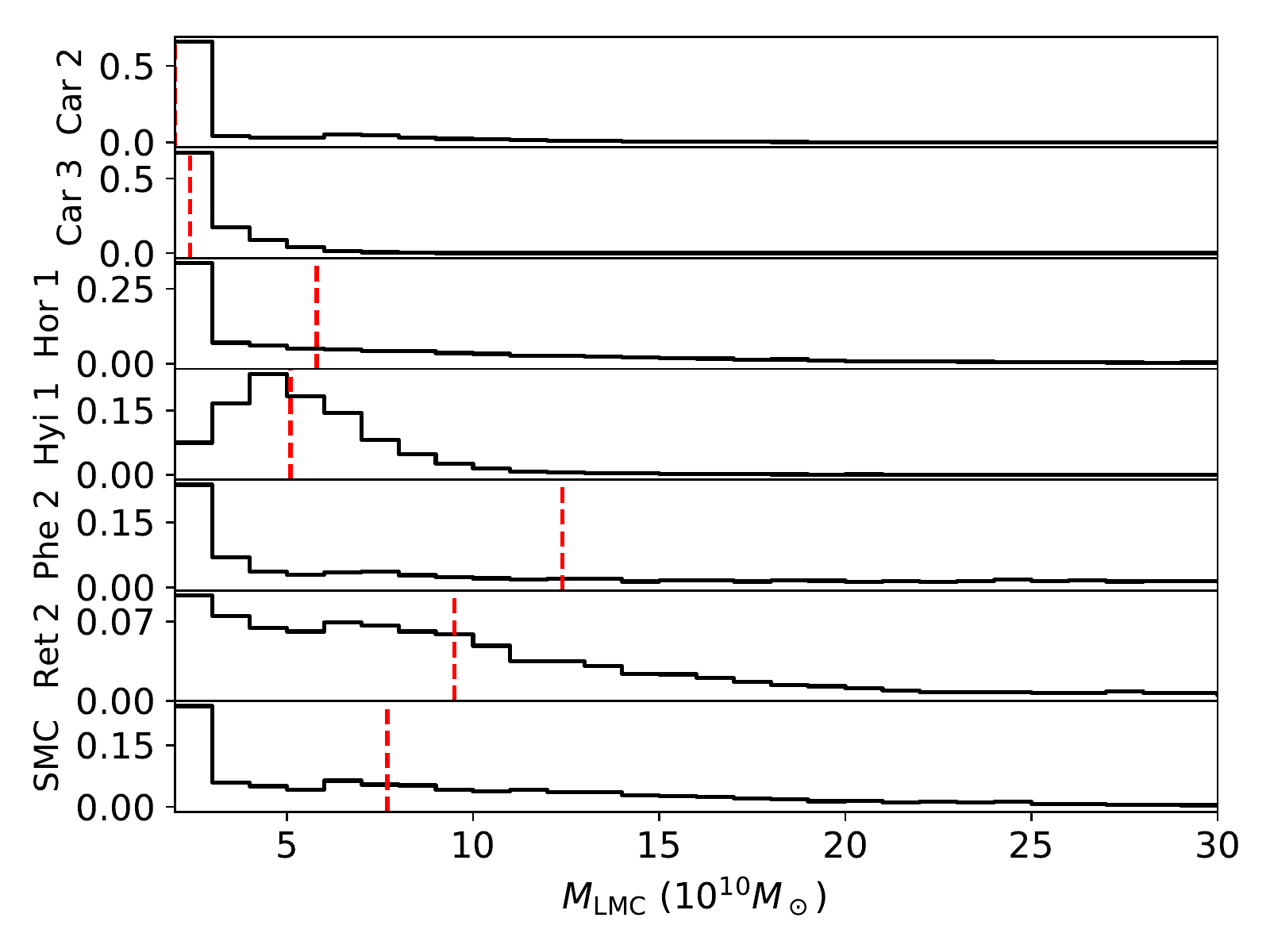}
\caption{Distribution of minimum mass needed to bind the satellites to the LMC. We show the six satellites which have a significant chance of being bound. Phe 2 requires the highest LMC mass in order to bind it, $12.4\times 10^{10} M_\odot$. } 
\label{fig:mlmc_bound}
\end{figure}

In \Figref{fig:mlmc_bound} we show the distribution of the minimum
mass needed to bind the satellites considered. Of all the satellites,
Ret 2 requires the highest LMC mass, $9.6\times 10^{10}
M_\odot$. Although this mass is larger than what has been measured in
the inner part of the LMC \citep{vandermarel_lmc}, it is significantly
less than the mass measured using the Orphan stream
\citep{orphan_modelling} or the mass measured using the nearby Hubble
flow and the timing argument with Andromeda
\citep{penarrubia_lmc}. Car 2 requires the lowest LMC mass in order to
bind it, consistent with its high probability of being bound even for
a low LMC mass (see \figref{fig:pbound}). The median LMC mass needed
to bind each satellite can be found in \Tabref{tab:sats}.

\subsection{Satellites without radial velocities}

This rewinding technique can also be used on satellites without radial
velocities. In order to do this, we sample the proper motions and
distance from their observed distributions, and uniformly sample the
radial velocities over the range -500 to 500 km/s. For each satellite,
we make 200,000 such samples and then bin the velocities with a width
of 10 km/s. For each velocity bins, we compute the probability of
being bound to a $1.5\times10^{11} M_\odot$ LMC, and then take the 
maximum probability across all of these bins. The
results are shown in \Tabref{tab:sats}. For the satellites with a
maximum probability above 0.4, we give the radial velocity range with
this probability. We see that there exist radial velocities for which
Eri 3 would have a high probability, 0.50, of being bound an LMC
satellite. We note that some of these satellites have large proper
motion errors (e.g. Hor 2 has errors of 0.42 mas/yr and 0.66 mas/yr)
which will naturally lead to a low probability of membership even for
genuine LMC's companions. Thus, the membership probability for these
satellites should be revisited once better proper motions are
available.

\subsection{Classical dwarfs} \label{sec:classical}

In order to assess whether any of the other classical dwarfs could
have fallen in with the LMC, we repeat our analysis for them with an
LMC mass of $1.5\times10^{11} M_\odot$ and
$2.5\times10^{11}M_\odot$. The results are shown in
\Tabref{tab:sats}. We find that for a $1.5\times10^{11} M_\odot$ LMC,
none of the classical satellites (excluding the SMC) have an
appreciable probability of belonging to the LMC. Increasing the LMC
mass to $2.5\times10^{11} M_\odot$, we find that Fornax has the
highest probability, 0.459, of having been bound to the LMC in the
past. Interestingly, \Figref{fig:vvel_rrel} shows that Carina is
closer to the LMC in phase space than Fornax. Despite this, Fornax has
a significantly higher probability of being bound to the LMC. This
shows that the present day binding energy can be misleading and one
needs to either rewind the satellites as done in this work, or model
their stripping from the LMC
\cite[e.g.][]{jethwa_lmc_sats,sales_etal_2017,kallivayalil_etal_2018}.

This analysis thus provides a useful counterpoint to those based on
the orbital plane alignment of the classical satellites
\citep[e.g.][]{pardy_etal_2019} which have suggested that both Carina
and Fornax may have originally been LMC satellites.

\subsection{Globular clusters}

In addition to dwarfs, we can also check if there are globular
clusters which were accreted along with the LMC. We repeat the same
analysis using the compendium of globular clusters with 3d positions
and velocities from \cite{vasiliev_GCs}. Note that this catalogue does
not include the 16 currently known globular clusters
\citep[e.g.][]{georgiev_etal_2010} which were unambiguously associated
to the LMC due to their proximity. We find that none of the globular
clusters considered are consistent with an LMC origin, and the cluster
with the highest probability of membership is Pal 14 which has a
modest probability, 0.135 (for an LMC mass of $1.5\times10^{11}
M_\odot$). It is unclear whether this lack of additional globular
clusters is surprising since the LMC already falls within the scatter
of the virial mass-globular cluster mass relation
\citep[e.g.][]{forbes_GCs}. However, this scatter is quite large for
galaxies in the mass range of the LMC, thus indicating that there
could be additional globular clusters. Furthermore, it is possible
that some of the ultra-faint dwarfs which we find to be LMC satellites
are in fact globular clusters \citep[e.g. Car 3,][]{car23_disc}.

\begin{table*}
\begin{centering}
\begin{tabular}{|c|c|c|c|c|c|c|}
Likely LMC satellites \\ \hline
Satellite & $p_{\rm LMC}(1.5\times10^{11} M_\odot)$ & $M_{\rm LMC}$ ($10^{10} M_\odot$) &  Source\\ \hline
Car 2 & 0.972 & 2.0  & $^a$ \\
Car 3 & 1.0 & 2.4 &$^a$  \\
Hor 1  & 0.753 & 5.8  & $^a$ \\
Hyi 1 & 0.996 & 5.1 & $^a$\\
Phe 2 & 0.532  &    12.4 & $^{b,c}$  \\
Ret 2 & 0.643 & 9.5 & $^a$ \\
SMC & 0.785 & 7.7 & $^{d,e,f}$\\  \hline \hline
\\ Inconsistent with LMC origin\\ \hline
Satellite & $p_{\rm LMC}(1.5\times10^{11} M_\odot)$ & $p_{\rm LMC}(2.5\times10^{11} M_\odot)$ &  Source\\ \hline
Ant 2 & 0.030 & 0.148 & $^g$ \\
Aqu 2 & 0.012 & 0.042 & $^h$ \\
Boo 1 & $<0.0001 $ &$<0.0001$ & $^a$\\
Boo 2 & 0.0007 & 0.045 & $^a$\\
Com & $< 0.0001$ & 0.007 & $^a$\\ 
Cra 2 & 0.007 & 0.149 & $^i$ \\
Dra 2 & 0.015 & 0.034 & $^a$ \\ 
Eri 2 & 0.001  & 0.002 & $^j$ \\
Gru 1 & 0.017 & 0.079 & $^j$ \\
Her 1 & 0.014      & 0.092 & $^i$     \\
Hya 2 & 0.045 & 0.121 & $^h$ \\
Seg 1 & 0.016 & 0.048 & $^a$ \\ 
Seg 2 & $<0.0001$ & $<0.0001$ & $^a$ \\
Tri 2 & 0.006 & 0.064 & $^a$ \\
Tuc 2 & 0.124 & 0.179 & $^a$ \\
Tuc 3 & 0.078 & 0.217 & $^a$ \\
UMa 1 & $< 0.0001$ &  $< 0.0001$ & $^a$ \\
UMa 2 & 0.0007 & 0.002 & $^a$ \\ 
Wil 1 & $<0.0001$ & 0.0004 & $^a$ \\ \hline
Carina &   $0.004$    & $0.125$ & $^{k,l}$  \\
Draco &    $0.087$       & $0.090$ &$^{k,l}$  \\ 
Fornax &   $0.128$  & $0.449$ & $^{k,l}$\\ 
Leo 1&   $0.132$  & $0.181$ & $^{k,l}$ \\
Leo 2  &  $0.073$  & $0.232$ & $^{k,l}$ \\
Sagittarius & $0.002$ & $0.031$ & $^{k,l}$ \\
Sculptor &  $<0.0001$ & $0.281$ & $^{k,l}$ \\
Sextans &    $<0.0001$ & $0.034$  & $^{k,l}$ \\
Ursa Minor &   $0.002$& $0.096$ &  $^{k,l}$ \\ \hline \hline
\\ Satellites with missing radial velocity \\ \hline
Satellite &  max$(p_{\rm LMC})$ & $v_r$ with $p_{\rm LMC} > 0.4$ & Source \\ \hline
Col 1 & 0.054 & $-$ & $^b$ \\
Eri 3 & 0.50 & [50,200] km/s &  $^b$ \\ 
Gru 2 & 0.15 & $-$ &  $^b$ \\ 
Hor 2 & 0.20 & $-$ & $^b$ \\ 
Pic 1 & 0.18 & $-$ &  $^b$ \\ 
Ret 3 & 0.0005 & $-$ &  $^b$ \\
Tuc 4 & 0.28 & $-$ &  $^b$ \\ 
 \hline \hline
\end{tabular}
\caption{Summary of results in this study. \textbf{Top table} shows the likely satellites of the LMC. The second column shows the probability of having been bound to the LMC assuming at LMC mass of $1.5\times10^{11} M_\odot$. The third column shows the median LMC mass needed to bind each satellite. \textbf{Middle table} shows satellites which have a low probability of being LMC satellites. For each satellite, we give the probability of having been bound to a $1.5\times10^{11} M_\odot$. \textbf{Bottom table} shows satellites which lack radial velocity information. The second column there shows the range of radial velocities for which the probability of being an LMC satellite is larger than 0.5 assuming an LMC mass of $1.5\times10^{11} M_\odot$. If the probability never rises above 0.5, the radial velocity is left blank. The data for the satellite come from a variety of sources: $^a$ is \protect\cite{simon_gaia_pms} and references therein, $^b$ is \protect\cite{pace_li_pms}, $^c$ is \protect\cite{fritz_rvs}, $^d$ is \protect\citep{smc_dist_2014}, $^e$ is \protect\citep{harris_zaritsky_2006}, $^f$ is \protect\citep{kallivayalil_etal_2013}, $^g$ is \protect\cite{antlia_disc}, $^h$ is \protect\cite{kallivayalil_etal_2018}, $^i$ is \protect\cite{fu_etal_2019}, $^j$ is \protect\cite{fritz_gaia_pms}, $^k$ is \protect\cite{gaia_pms}, $^l$ is \protect\cite{mcconnachie_2012}. }
\label{tab:sats}
\end{centering}
\end{table*}

\section{Testing with a cosmological simulation} \label{sec:cosmo_tests}

In this section we test this method of identifying LMC group members in a cosmological zoom-in simulation of a Milky Way-like halo with an LMC analogue at the present. This simulation is described in \cite{jethwa_upper_lower} and is evolved with the $N$-body part of \textsc{gadget-3} which is an updated version of \textsc{gadget-2} \citep{springel_2005}.

In order to compare with the LMC, we choose the snapshot shortly after the pericenter of the LMC analogue. This occurs at a scale factor of 0.96 which corresponds to a lookback time of 0.5 Gyr. At this time, the Milky Way-analogue has a virial mass of $8.77\times10^{11} M_\odot$, a scale radius of $20.5$ kpc, and a concentration of $12.1$. In the rewinding procedure, we use these values for the Milky Way potential at all times. At the same scale factor, the LMC analogue has a mass of $7.01\times10^{10}M_\odot$ and a peak mass of $7.96\times10^{10} M_\odot$. For consistency, we treat the LMC as a Hernquist profile which matches the circular velocity, $69$ km/s, of the LMC-analogue at a given radius which we choose to be 5 kpc. 

We select the 20 closest satellites to the LMC in distance. 14 of
these were originally LMC satellites given that they were within the
LMC's virial radius before the LMC was accreted onto the Milky
Way. For each satellite, we make mock observations from the location
of the Sun. We include an error in the distance (6.4\% distance
error), radial velocity (1.15 km/s), and proper motions (0.1015, 0.098
mas/yr in $\mu_\alpha^*,\mu_\delta$ respectively) based on the median
errors in our sample of dwarfs. This is also done for the LMC-analogue
with errors from the LMC observations in the literature.

\begin{figure}
\centering
\includegraphics[width=0.45\textwidth]{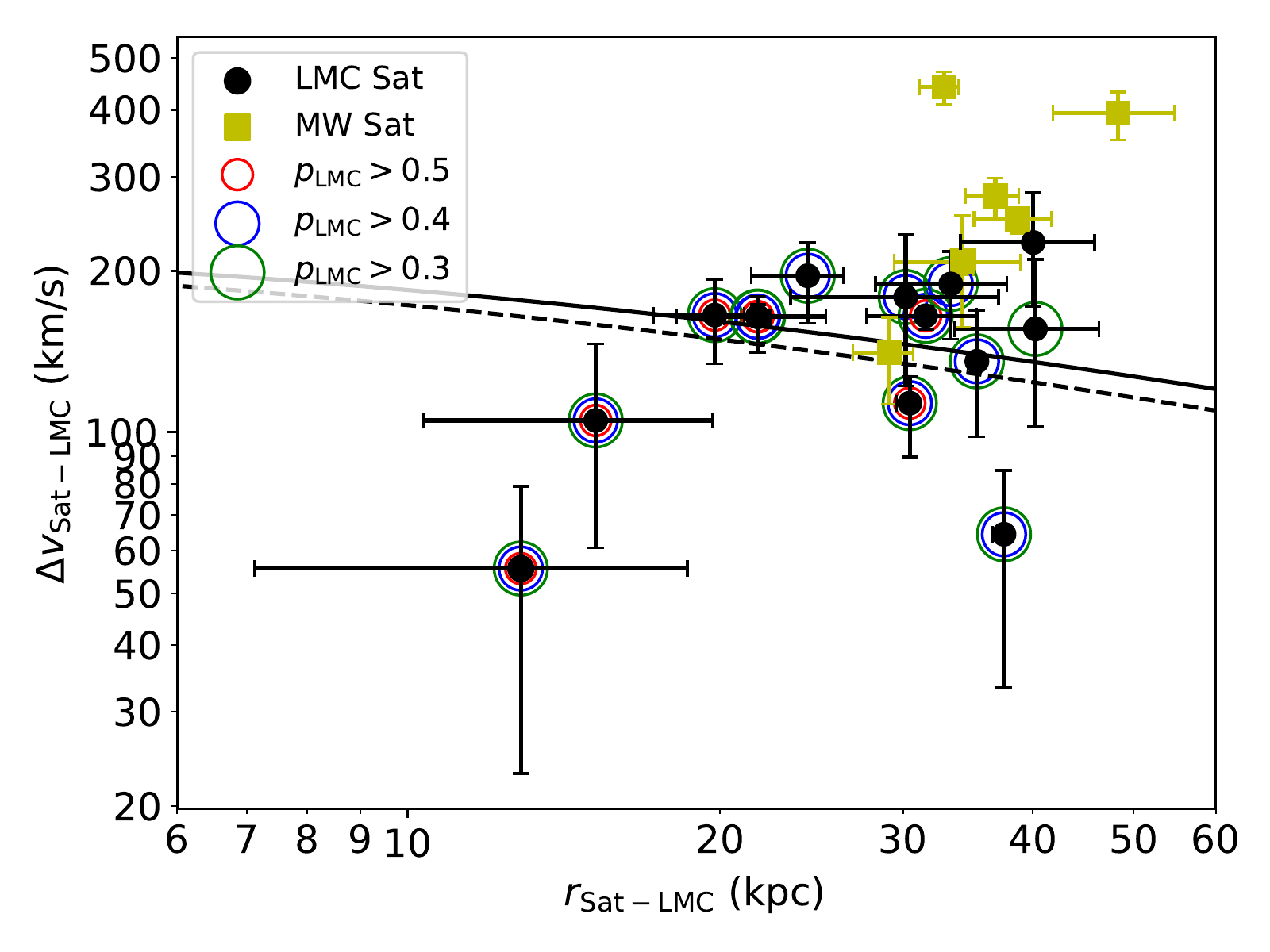}
\caption{Relative position and velocity between the LMC analogue and dwarf galaxy analogue in our cosmological simulation. The solid line shows the escape velocity curve when the LMC-analogue achieved its peak mass and the dashed line shows the escape velocity curve at the present. The colored rings show the probability of having been bound when using the peak mass of the LMC. Interestingly, many of the LMC's satellites are energetically unbound with respect to the LMC-analogue.  Thus, using only the present-day position and velocity would lead to an overestimate of the LMC's mass. We also see that this method does a good job separating the Milky Way satellite closest to the LMC. This satellite only has a $28.5\%$ chance of belonging to the LMC despite appearing to be energetically bound. } 
\label{fig:mlmc_bound_mock_MW}
\end{figure}

The position and velocity of these satellites relative to the LMC is shown in \Figref{fig:mlmc_bound_mock_MW}. For reference, the escape velocity curves of the LMC-analogue given its peak mass (present-day mass) is shown as a solid-black (dashed-black) line. Interestingly, many of the LMC satellites are energetically unbound at the present day. This difference is due to the tidal field of the Milky Way-analogue. This highlights the fact that using the present-day escape velocity could lead us to miss a large fraction of the LMC satellites.

Given the mock observations made of these satellites, the approach is
identical to what is described in \Secref{sec:mw_analysis}. First, we
calculate the probability of being bound to the LMC-analogue as a
function of its mass and then compute the distribution of masses
needed to bind each satellite. As a test of the method, we compute the
probability of being bound to the LMC-analogue using the peak mass of
the LMC ($7.96\times10^{11}M_\odot$) and encircle these with a red,
blue, green ring in \Figref{fig:mlmc_bound_mock_MW} if the probability
is greater than 0.5, 0.4, 0.3 respectively. This figure shows that the
method presented in this work can distinguish between Milky Way and
LMC satellites close to the LMC with a high fidelity. Indeed, the
Milky Way satellite with the highest chance of belonging to the LMC
only has a 28.5\% chance of being an LMC satellite. As expected, the
technique performs better for satellites closer to the LMC in phase
space, but it is also capable of correctly classifying LMC satellites
which are unbound at the present. Finally, this figure also advocates
that even satellites with a relatively modest probability of being an
LMC satellite ($p_{\rm LMC} > 0.3$) should strongly be considered.

\begin{figure}
\centering
\includegraphics[width=0.45\textwidth]{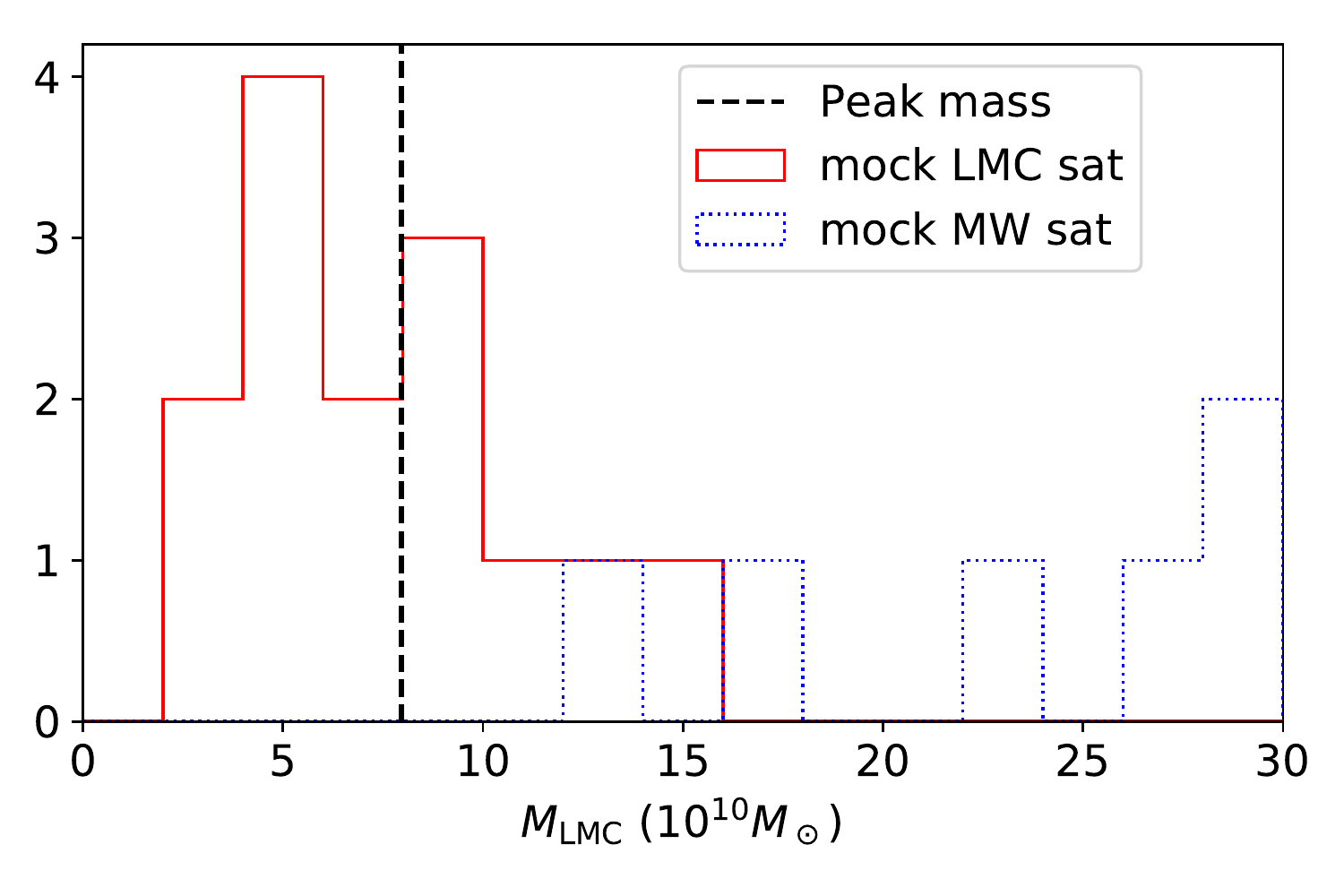}
\caption{Minimum mass needed to bind satellites to the LMC analogue in the cosmological simulation. The red histogram shows the median mass needed for LMC satellites while dashed-blue histogram shows the minimum mass needed for Milky Way satellites. The vertical dashed-black line shows the peak mass of the LMC in the simulation. While 8 of the true LMC satellites require a mass less than the LMC's peak mass, 6 require a higher mass. Thus, with the current errors, it is possible that the LMC mass will be overestimated with this technique. Furthermore, if a Milky Way satellite is misclassified as an LMC satellite, this will also lead to an overestimate of the LMC mass.} 
\label{fig:mlmc_bound_mass}
\end{figure}

We also compute the distribution of minimum LMC masses needed to bind
each mock satellite as in \Secref{sec:mw_analysis}. This distribution
is shown in \Figref{fig:mlmc_bound_mass} with the satellites of the
LMC analogue shown in solid-red and the satellites of the Milky Way
analogue shown in dotted-blue. We see that 6 out of 14 satellites of
the LMC analogue have median binding masses larger than the peak mass
of the LMC analogue. This shows that the LMC mass estimate in
\Secref{sec:mw_analysis} may be overestimated. Applying the same
technique on the satellites of the Milky Way analogue gives
significantly higher LMC masses since the Cloud must strip them from
the Milky Way during the rewinding process.

Finally, our mock dwarf catalogue allows us to test how this satellite
classification will look with reduced errors from the future
observations. In order to assess this, we repeat the analysis assuming
errors which are half as large in distance, radial velocity, and
proper motion. With these reduced errors, the Milky Way satellite with
the highest probability of belonging to the LMC given the LMC's peak
mass only has $p_{\rm LMC} = 0.097$. Furthermore, all but 3 LMC
satellites have $p_{\rm LMC} > 0.5$. Thus, as the data quality
improves in the future, this method will become more precise in
distinguishing LMC satellites from those of the Milky Way. This will
also improve the mass estimate of the LMC.

\section{Discussion} \label{sec:discussion}

\subsection{Comparison with previous works}

Without proper motions (and for many objects without the radial
velocity), \cite{jethwa_lmc_sats} made predictions as to which dwarfs
were accreted with the LMC. Of their 7 highest confidence candidates
($p_{\rm LMC} > 0.7$), we confirm that Hor 1, Ret 2, and Phe 2 are
likely LMC satellites. We also find that there are radial velocities
for which Hor 2 and Tuc 4 have modest probabilities (0.2, 0.28
respectively) of being an LMC satellite.  However, we find that Tuc 2
is very unlikely to be an LMC satellite. Among their 5 medium
probability candidates ($0.5 < p_{\rm LMC} < 0.7$), we rule out Tuc 3
and Gru 1. The other 3 (Gru 2, Pic 1, Ret 3) are missing radial
velocities although we note that the maximum probability of being
bound for Ret 3 is 0.0005 and thus it is very unlikely to be an LMC
satellite.

Similarly, \cite{sales_etal_2017} predicted the properties of LMC satellites
with known satellites and found that Hor 1 agreed with the properties known
at that time (i.e. without proper motions). In addition, they found Hor 2, Eri 3, Ret 3, Tuc 4, Tuc 5, and Phe 2  had on-sky positions and distances consistent with a Magellanic origin. Of these, we confirm Phe 2 is likely an LMC satellite. We also find that there are radial
velocities for which Eri 3 would likely be an LMC satellite. Their 
remaining satellites still do not have complete measurements of their kinematics.
However, as mentioned above, we find Ret 3 is very unlikely to be an LMC satellite.

Finally, our results can also be compared with those of
\cite{kallivayalil_etal_2018}. Using a different technique based on a
single cosmological simulation, they identified Car 2, Car 3, Hor 1,
and Hyi 1 as likely LMC satellites, in agreement with this work. They
also identified Ret 2, Tuc 2, and Gru 1 as inconsistent with the LMC
at the $3\sigma$ level in velocity. However, our analysis suggests
that Ret 2 is likely to be an LMC satellite. This discrepancy may be
due to the modest LMC mass of $3.6\times 10^{10} M_\odot$ used in
\cite{kallivayalil_etal_2018}. Furthermore, they suggest that Hya 2
and Dra 2 could be LMC satellites although these are essentially ruled
out as LMC satellites in this work. Finally, they also argued that Phe
2 could be an LMC satellite based on their prediction of its proper
motion and the existence of an overdensity of stars with that proper
motion in \textit{Gaia} DR2. Our analysis which makes use of the
observed proper motion and radial velocity from \cite{fritz_rvs} finds
that Phe 2 is a likely LMC satellite.

\subsection{Structural properties of dwarfs}

\begin{figure}
\centering
\includegraphics[width=0.45\textwidth]{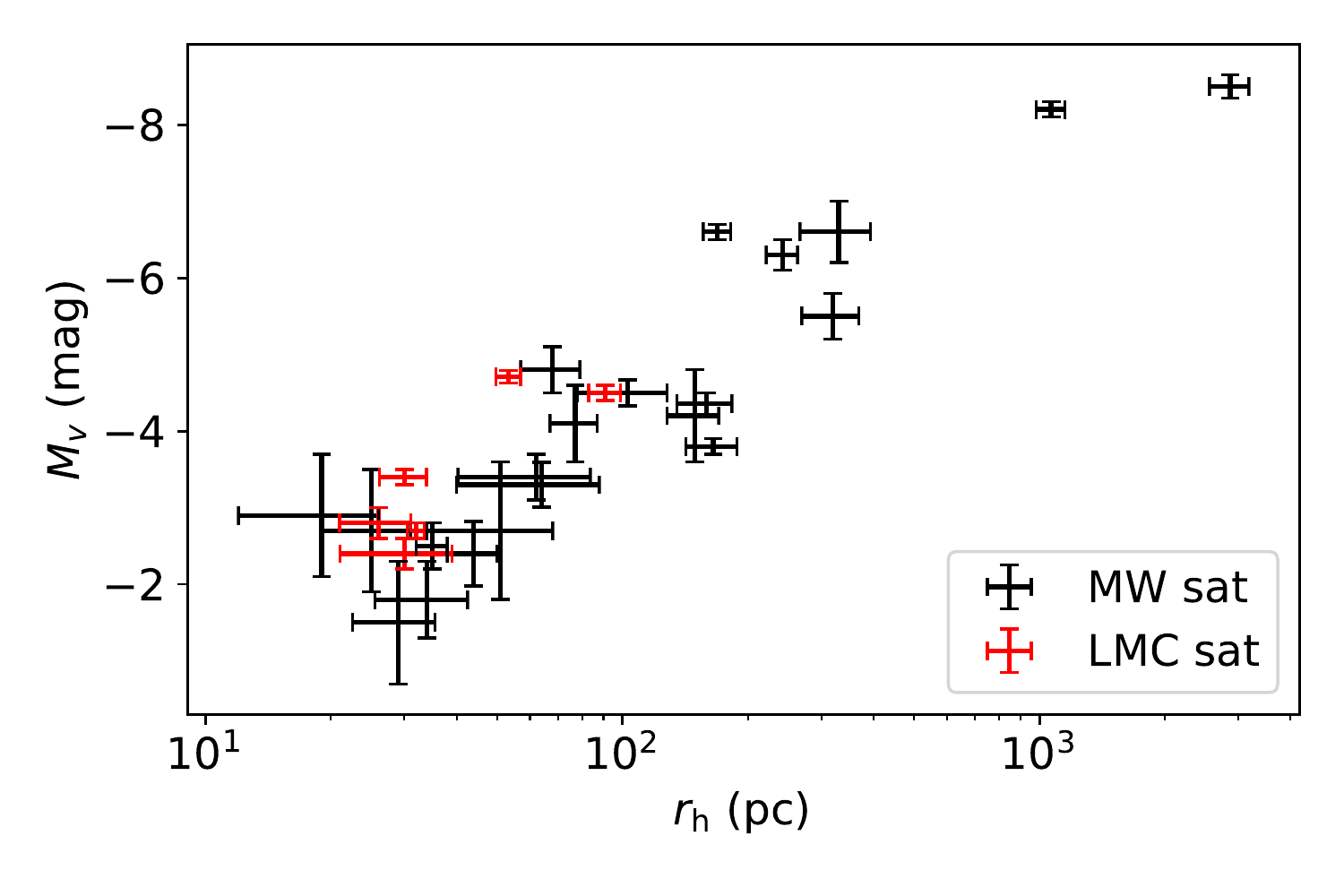}
\caption{Half-light radius versus magnitude for our sample of satellites. The likely LMC satellites are shown in red while the other satellites inconsistent with an LMC origin are shown in black. We also include Col 1 and Ret 3 as not having an LMC origin since there are no radial velocities for which they would have an appreciable probability of being an LMC satellite. The properties of these dwarfs come from \protect\cite{drlica_wagner_etal_2016,pic2_disc,koposov_des_sats,hydrus_disc,tri2_disc,dra2sag2lev3_disc,hya2_disc,mcconnachie_2012,cra2_disc,car23_disc,antlia_disc}. }
\label{fig:mv_vs_rhalf}
\end{figure}

Now that we can distinguish which satellites belong to the Milky Way
and the LMC, we can contrast their structural properties. In
\Figref{fig:mv_vs_rhalf} we compare the half-light radius versus
magnitude for objects with $6d$ data; here the LMC satellites are
shown in red and the Milky Way satellites shown in black. At a fixed
magnitude, the LMC satellites appear to be smaller. This could be a
sign that the tidal radii of the LMC satellites are smaller since they
orbit a less massive host compared to the Milky Way satellites. This
is also interesting in the context of similar comparisons between
Milky Way satellites and those of M31 where it was found that the M31
satellites were larger by a factor of 2
\citep{mcconnachie_irwin_2006}. However, a subsequent and more
detailed analysis concluded that this difference was not statistically
significant \citep{brasseur_etal_2011}.

\subsection{Satellite planes around the LMC}

\cite{jethwa_lmc_sats} found 7 satellites (Gru 1, Gru 2, Phe 2, Tuc 2,
Tuc 3, Tuc 4, Tuc 5) reside in a thin plane around the LMC. This is
especially interesting given the planes of satellites around the Milky
Way \citep[e.g.][]{lynden-bell_1976,pawlowski_vpos_2012}, Andromeda
\citep{ibata_M31_plane}, and Centaurus A
\citep{mueller_etal_2018}. While they found this plane was
significant, they also noted that their models showed that Gru 1, Gru
2, and Tuc 3 were likely to be MW satellites. From the results of this
work, we confirm that Gru 1 and Tuc 3 are not associated with the
LMC. Gru 2 is also not associated with the LMC since there is no
radial velocity for which it has a significant LMC membership
probability. Furthermore, we find that Tuc 2 is also not associated
with the LMC. Of the remaining satellites, we confirm that Phe 2 is an
LMC satellite, that Tuc 4 has a chance of being an LMC satellite. We
did not investigate Tuc 5 due to the lack of proper
motions. Therefore, at most 3 out of the original 7 satellites are
associated with the LMC which makes the plane less significant.

\begin{figure}
\centering
\includegraphics[width=0.45\textwidth]{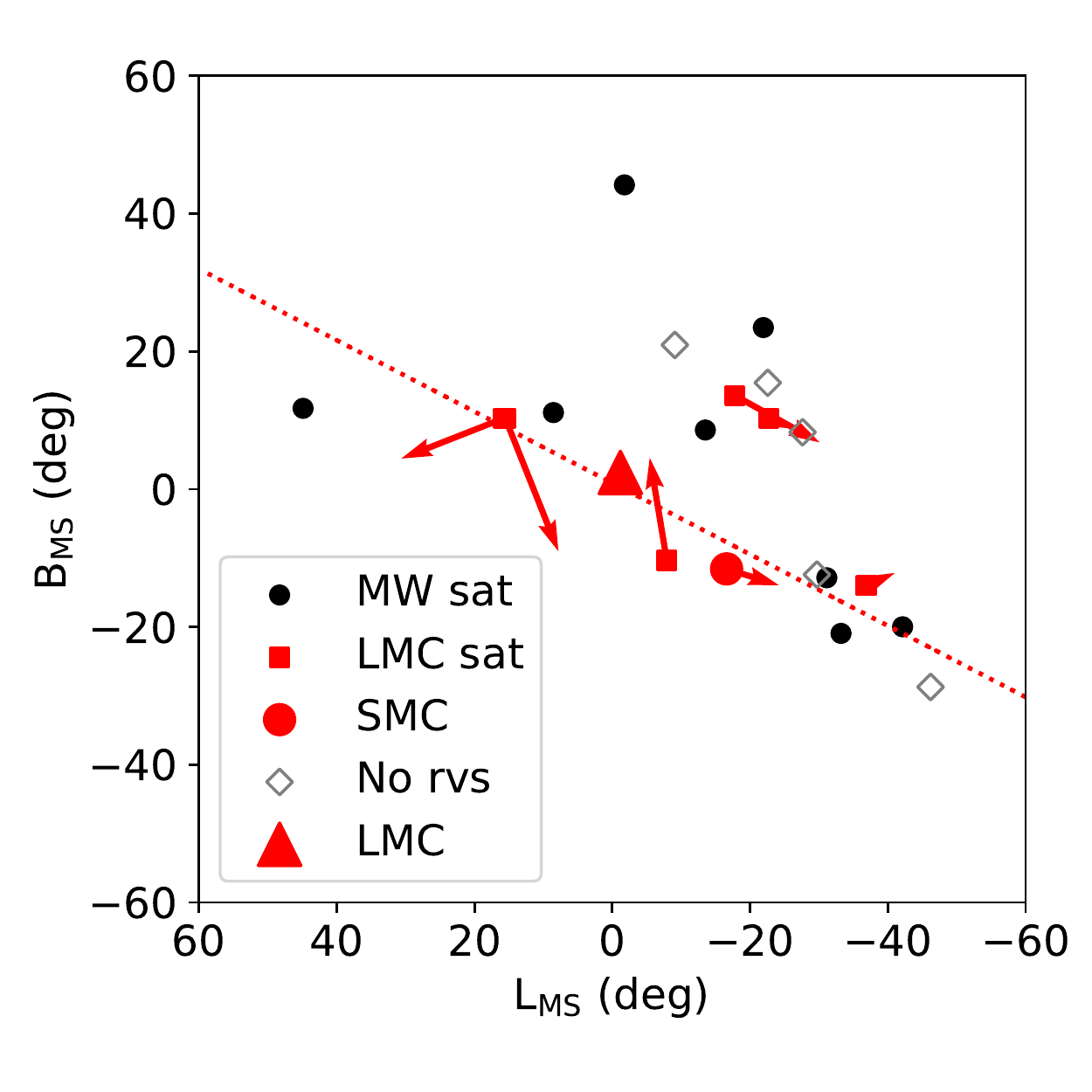}
\caption{Possible plane of satellites around the LMC. The red triangle shows the LMC, the red circle shows the SMC, and the red squares show high probability LMC satellites identified in this work. The black circles show the satellites inconsistent with an LMC origin and the grey diamonds show satellites without radial velocities for which we did determine membership. Arrows show proper motions relative to the LMC as described in the text. The two satellites without radial velocities close to the plane are Tuc 4 and Gru 2.} 
\label{fig:plane}
\end{figure}

In \Figref{fig:plane} we show the satellites near the LMC along with
their classification from this work. While the plane from
\cite{jethwa_lmc_sats} is now less significant, we find that the
majority (5 out of 7) of the known LMC satellites do reside in a thin
plane. In order to check whether the kinematics of these satellites
are consistent with a long-lived plane, we show their proper motions
of relative to the LMC. This is done by taking the 3d velocity of each
satellite, subtracting the 3d velocity of the LMC, and computing the
resulting proper motion as viewed from the Sun. Since many of the LMC
satellites have substantial velocity components perpendicular to the
plane, we conclude that the plane is likely a chance alignment.

\subsection{Change to satellite orbits}

Since we follow the orbits of the satellites over the past 5 Gyr, we can study how these change due to the inclusion of the LMC. This is particularly useful for understanding which satellites should be tidally affected by the Milky Way. For this comparison, we use the orbits computed in \Secref{sec:lmc_prob} for an LMC mass of $2.5\times10^{11} M_\odot$, as well as orbits computed without the influence of the LMC. For each satellite, we compute the median pericenter with respect to the Milky Way. In \Tabref{tab:orbits} we show the satellites whose median pericenters change by 25\% or more. Note that in this search we ignore the 7 satellites we have associated with the LMC.

Interestingly, both Ant 2 and Cra 2 have significantly smaller
pericenters in the presence of the LMC. Both of these satellites are
quite diffuse and strong tidal disruption has been suggested as being
responsible for this
\citep[e.g.][]{sanders_etal_2018,antlia_disc}. This shows that it is
crucial to model these systems in the presence of the LMC if we want
to understand their genesis.

\begin{table}
\begin{centering}
\begin{tabular}{|c|c|c|}
Satellite & Pericenter inc. LMC & Change to pericenter  \\
\hline Ant 2 & 25.7$^{+9.3}_{-7.6}$ kpc & 0.63 \\
Cra 2 & 19.7$^{+11.3}_{-7.6}$ kpc & 0.41 \\
Her 1 & 72.3$^{+22.0}_{-27.6}$ kpc & 1.25 \\
Tuc 3 & 2.1$^{+1.8}_{-1.6}$ kpc & 0.75 \\
\hline Carina & 73.3$^{+25.3}_{-13.4}$ kpc & 0.69 \\
Draco & 73.1$^{+7.9}_{-10.1}$ kpc & 1.96 \\
Ursa Minor & 71.3$^{+6.7}_{-6.4}$ kpc & 1.66 \\
 \hline 
\end{tabular}
\caption{Change to pericenter of satellite orbit with respect to the Milky Way. The middle column gives the median pericenter, with $1 \sigma$ scatter, of the satellite with respect to the Milky Way in the presence of a $2.5\times10^{11} M_\odot$ LMC. The third column shows the fractional change of this median pericenter from satellite orbits without the influence of the LMC.}
\label{tab:orbits}
\end{centering}
\end{table}

\subsection{Expected number of satellites around the LMC}

\cite{jethwa_lmc_sats} predicted that the number of LMC satellites in
the magnitude range $-7 < M_V < -1$ was $70^{+30}_{-40}$. This
calculation was based on matching the observed luminosity function and
used membership probabilities calculated without any proper motion
information. Now that we have an improved determination of the
satellites' LMC membership, we can update this calculation. This is
done by taking the probabilities of being an LMC satellite from
\cite{jethwa_lmc_sats} and replacing them with 1 or 0 depending on how
we have classified them in this work. If the satellites do not have $6d$
information (i.e. missing proper motions or radial velocities), we
keep the probability as determined by \cite{jethwa_lmc_sats}. We then
compute the expected number of LMC satellites with these updated
probabilities, compare this with the expectation value in
\cite{jethwa_lmc_sats}, and scale the total number of satellites by
this ratio. This reduces the expected number of satellites slightly to
60. Thus we conclude that there is a large number of LMC satellites
still to be found given that we have only identified 7 in this work.

\section{Conclusions} \label{sec:conclusion}

This work confirms the LMC group infall found in
\cite{kallivayalil_etal_2018} with an independent technique and
extends the number of LMC satellites. In addition to the SMC, we
identify six dwarf galaxies as being highly likely bound to the Cloud
previously: Car 2, Car 3, Hor 1, Hyi 1, Phe 2, and Ret 2. Furthermore,
we find that there exist radial velocities for which Eri 3 has a
significant probability of being an LMC satellite ($p_{\rm LMC} >
0.4$). Having tested our method on cosmological simulations, we
believe that dwarfs with a modest probability of being LMC satellites
(i.e. Hor 2 and Tuc 4) should be reconsidered once additional (better)
data is available. Tests with cosmological simulations also showed
that as the observational errors are reduced, the certainty with which
we can classify satellites as belonging to the LMC will improve.

Assuming that these 7 satellites identified here do belong to the LMC,
we estimate that the Cloud's mass should be $>1.24\times10^{11}
M_\odot$. This estimate is tested with a cosmological simulation where
we find that given the current observational errors, this lower bound
may be overestimated. As the observational errors are reduced, this
lower bound will become more accurate and thus it should be revisited
once better proper motions, e.g. from \textit{Gaia} DR3, are
available.

We also explore the orbits of the Milky Way satellites and find that
several satellites have a large change in their orbit once the LMC is
included. Most interestingly, in the presence of the LMC, both Ant 2
and Cra 2 are on orbits with small(er) pericentric passages with
respect to the Milky Way. It is possible that tidal disruption on
these adjusted orbits can naturally explain their diffuse properties
\citep[as proposed in][]{sanders_etal_2018,antlia_disc} which shows
that it is crucial to include the LMC when studying the evolution of
Milky Way satellites.

The ability to distinguish LMC satellites from that of the Milky Way
opens up the prospects of better understanding the ultra-faint dwarf
satellites of both the Milky Way and the Cloud. Previous analyses of
the Milky Way satellites have avoided the dwarfs recently discovered
in the DES data due to the possible contamination by the LMC
\citep[e.g.][]{jethwa_upper_lower}. Since the ultra-faint dwarfs in
DES are also some of the faintest satellites detected, the constraints
these studies placed on the nature of dark matter can now be improved.

\section*{Acknowledgements}

We thank Ting Li for valuable discussions. This work has made use of data from the European Space Agency (ESA) mission {\it Gaia}(\url{https://www.cosmos.esa.int/gaia}), processed by the {\it Gaia} Data Processing and Analysis Consortium (DPAC, \url{https://www.cosmos.esa.int/web/gaia/dpac/consortium}). Funding for the DPAC has been provided by national institutions, in particular the institutions participating in the {\it Gaia} Multilateral Agreement.

This research made use of \textsc{ipython} \citep{IPython}, python packages \textsc{numpy} \citep{numpy}, \textsc{matplotlib} \citep{matplotlib}, and \textsc{scipy} \citep{scipy}. This research also made use of Astropy,\footnote{http://www.astropy.org} a community-developed core Python package for Astronomy \citep{astropy:2013, astropy:2018}.

\bibliographystyle{mn2e_long}
\bibliography{citations_lmc}

\end{document}